\documentclass[a4paper, amsfonts, amssymb, amsmath, print, showkeys, nofootinbib, twocolumn,twoside,floatfix,prb,superscriptaddress]{revtex4-2}
\usepackage[english]{babel}
\usepackage[utf8]{inputenc}
\usepackage{soul}
\usepackage{xcolor}
\usepackage[colorinlistoftodos, color=green!40, prependcaption]{todonotes}
\usepackage[pdftex, pdftitle={Article}, pdfauthor={Author}]{hyperref} 
\usepackage[normalem]{ulem}
\usepackage{xcolor}
\usepackage{amsthm}
\usepackage{mathtools}
\usepackage{physics}
\usepackage{xcolor}
\usepackage{graphicx}
\usepackage[left=23mm,right=13mm,top=35mm,columnsep=15pt]{geometry} 
\usepackage{adjustbox}
\usepackage{placeins}
\usepackage[T1]{fontenc}
\usepackage{lipsum}
\usepackage{csquotes}
\usepackage{romannum}
\usepackage{tikz}
\usepackage{miller}
\usepackage{romannum}
\usepackage{float}
\usepackage{placeins}
\usepackage{svg}
\usepackage{comment}

\newcommand{\B}{\(\mu_0\textit{H}\)}
\newcommand{\qh}{\(\text{q}_\text{h }\)}
\newcommand{\hcone}{\(\text{H}_\text{c1 }\)}
\newcommand{\hctwo}{\(\text{H}_\text{c1 }\)}
\newcommand{\tc}{\(\text{T}_\text{c }\)}

\begin{document}
\title{Observation of distorted tilted conical phase at the surface of a bulk chiral magnet with resonant elastic x-ray scattering}

\author{S. Mehboodi}
    \affiliation{Physik-Department, Technische Universität München, D-85748 Garching, Germany}
    \affiliation{Munich Center for Quantum Science and Technology (MCQST), Munich, Germany}
\author{V. Ukleev}
    \affiliation{Helmholtz-Zentrum Berlin für Materialien und Energie, Berlin, Germany}
\author{C. Luo}
    \affiliation{Helmholtz-Zentrum Berlin für Materialien und Energie, Berlin, Germany}
\author{R. Abrudan}
    \affiliation{Helmholtz-Zentrum Berlin für Materialien und Energie, Berlin, Germany}
\author{F. Radu}
    \affiliation{Helmholtz-Zentrum Berlin für Materialien und Energie, Berlin, Germany}
\author{C. H. Back}
    \affiliation{Physik-Department, Technische Universität München, D-85748 Garching, Germany}
    \affiliation{Munich Center for Quantum Science and Technology (MCQST), Munich, Germany}
     \affiliation{Center for Quantum Engineering (ZQE), Technical University Munich, D-85748 Garching, Germany}
\author{A. Aqeel}
    \affiliation{Physik-Department, Technische Universität München, D-85748 Garching, Germany}
    \affiliation{Munich Center for Quantum Science and Technology (MCQST), Munich, Germany}
    \affiliation{Institute of Physics, University of Augsburg, Augsburg 86159, Germany}


\begin{abstract}
We report on various magnetic configurations including spirals and skyrmions at the surface of the magnetic insulator Cu$_2$OSeO$_3$ at low temperatures with a magnetic field applied along ⟨100⟩ using resonant elastic X-ray scattering (REXS). We observe a well-ordered surface state referred to as a distorted tilted conical spiral (TC) phase over a wide range of magnetic fields. The distorted TC phase shows characteristic higher harmonic magnetic satellites in the REXS reciprocal space maps. Skyrmions emerge following static magnetic field cycling and appear to coexist with the distorted TC phase. Our results indicate that this phase represents a distinct and stable surface state that does not disappear with field cycling and persists until the field strength is increased sufficiently to create the field-polarized state.
\end{abstract}

\keywords{Resonant elastic X-ray scattering, magnetic structure, distorted tilted conical, multidomain low-temperature skyrmion}

\maketitle

\section{Introduction} \label{sec:intro}

In recent years, cubic chiral magnets have gained significant interest for their ability to host unique chiral magnetic spin configurations like helical spirals, chiral soliton lattices~\cite{okamura2017emergence,Brearton2023}, skyrmions~\cite{muhlbauer2009skyrmion,tonomura2012real,seki2012}, and screw dislocations~\cite{Azhar2022}. These noncollinear spin configurations are largely driven by the antisymmetric Dzyaloshinskii-Moriya interaction intrinsic in these chiral magnets, offering new avenues for emerging computing technologies \cite{song2020skyrmion,li2023magnetic,lee2024task}. The topologically nontrivial skyrmion textures were initially observed in the cubic chiral magnet MnSi using neutron scattering techniques \cite{muhlbauer2009skyrmion} and subsequently in Cu$_2$OSeO$_3$ \cite{seki2012}. The latter is a unique member of the cubic chiral magnet family because it is an insulator with a wide band gap of 2.4 eV~\cite{Versteeg2016} providing the opportunity to control the skyrmion lattice phase by electric fields~\cite{huang2018situ,white2014electric}. In addition, the low Gilbert damping \cite{stasinopoulos2017low,weiler2017helimagnon} observed in Cu$_2$OSeO$_3$ makes it promising for potential applications in magnonic devices \cite{barman20212021}. Cu$_2$OSeO$_3$ hosts two independent skyrmion pockets at different magnetic field-temperature regions of the phase diagram~\cite{chacon2018observation,bannenberg2019multiple,aqeel2021microwave}. It has been proposed, that the competition between cubic magneto-crystalline and exchange anisotropies in Cu$_2$OSeO$_3$ at low temperatures \cite{baral2023direct} leads to the stabilization of novel noncollinear magnetic phases, such as the tilted conical state and the low-temperature skyrmion lattice phase. Notably, these phases are only observed when the magnetic field is applied along the crystallographic $\hkl<100>$ directions \cite{qian2018new,halder2018thermodynamic,chacon2018observation,bannenberg2019multiple}.

\begin{figure}[hb]
    \centering
    \includegraphics[width=0.46\textwidth]{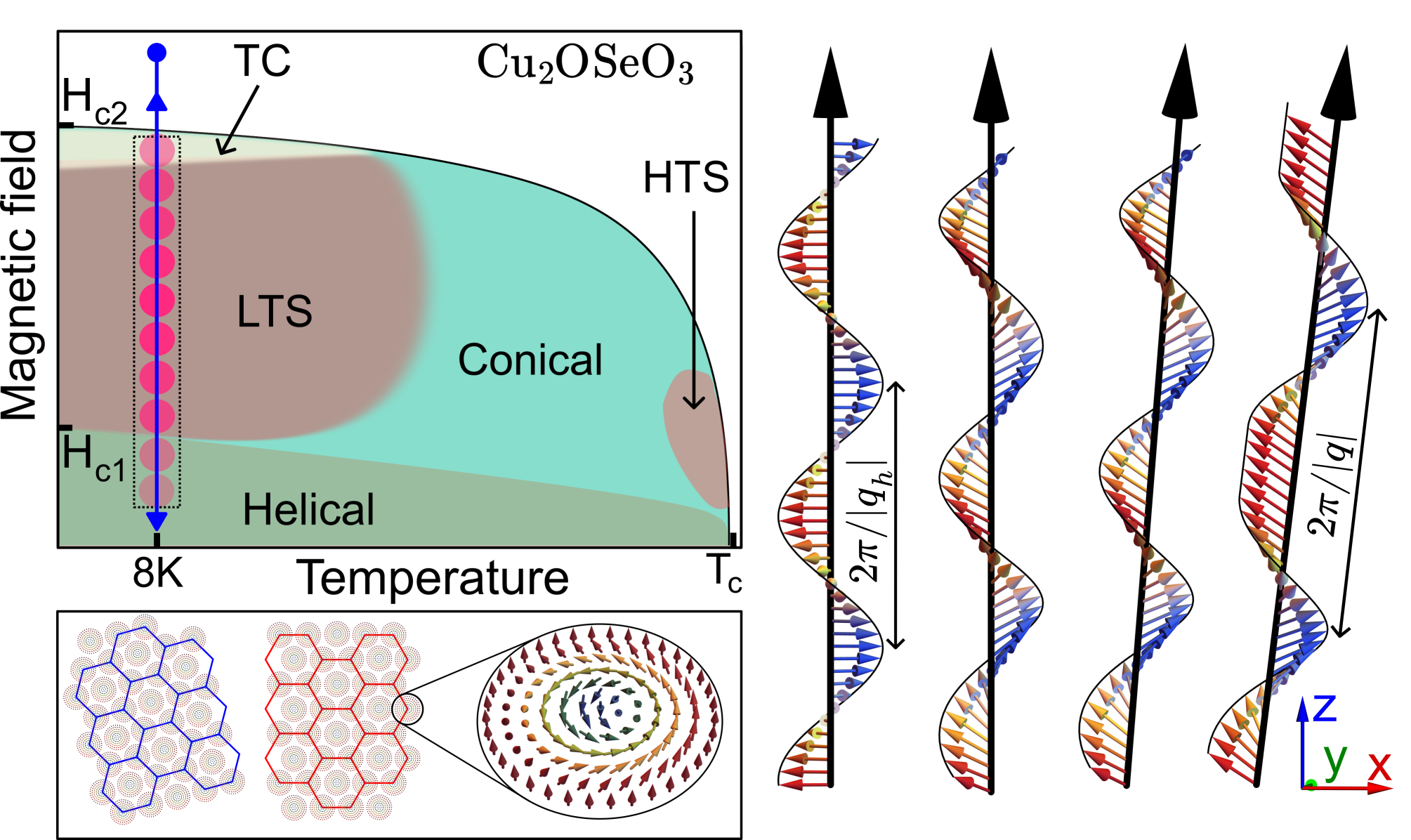}
    \put(-235,130){\text{(a)}}
    \put(-235,40){\text{(b)}}
    \put(-100,0){(\text{c})}
    \put(-75,0){(\text{d})}
    \put(-50,0){(\text{e})}
    \put(-30,0){(\text{f})}
    \caption{a) Illustration of the magnetic phase diagram for Cu$_2$OSeO$_3$ for magnetic field applied along $\hkl<001>$. HTS, LTS, and TC indicate high-temperature skyrmion, low-temperature skyrmion, and titled conical spiral phases. The rectangle highlights the field region where distorted TC spirals (magenta circles) are observed, at $T\approx$ 8 K. The blue arrows indicate the magnetic field sweep direction. b) Shows a schematic representation of two differently oriented hexagonal skyrmion lattices (blue and red) featuring Bloch-type skyrmions. c),d),e), and f) show the magnetic configurations of helical, conical, TC, and distorted TC spirals, respectively.}
    \label{csophase}
\end{figure}

\begin{figure*}[t!]
    \centering
    \includegraphics[]{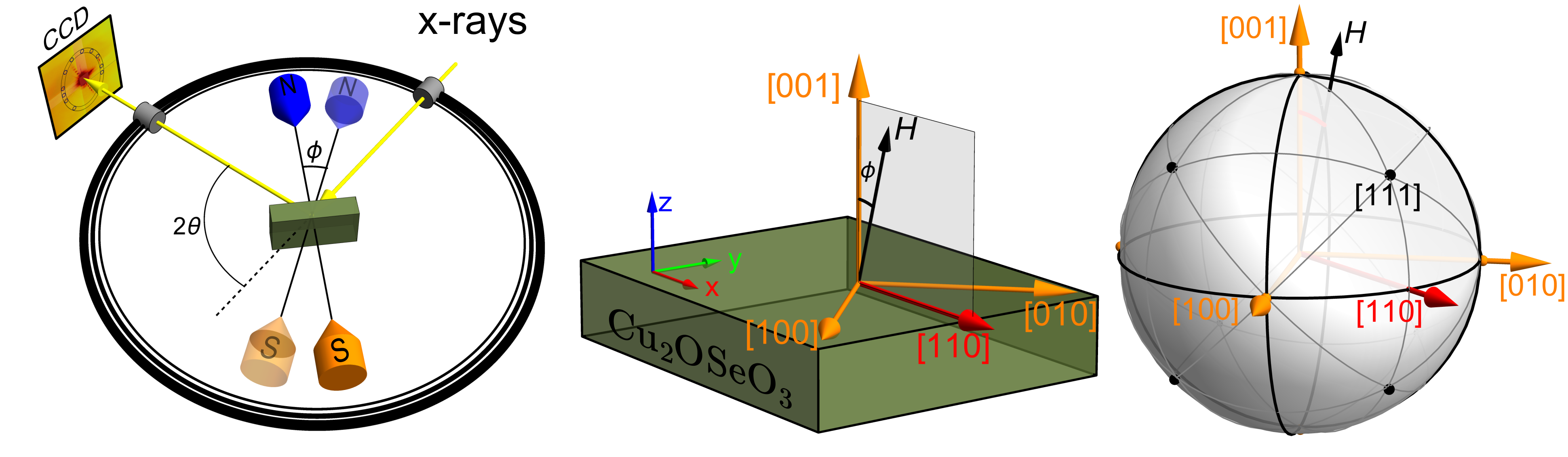}
    \put(-420,135){\text{(a)}}
    \put(-300,135){\text{(b)}}
    \put(-140,135){\text{(c)}}
    \vspace{0.1cm}
    \caption{(a) Schematic representation of the experimental setup. The X-ray beam is scattered off the sample satisfying the 2$\theta=$96.5$^\circ$ diffraction condition and is captured by a CCD camera. $\phi$ represents the rotational angle of magnetic field w.r.t to the sample normal ($\hkl<001>$) b) and c) highlight the coordinate system of field orientation relative to crystallographic directions of the sample.}
    \label{measgeo}
\end{figure*}

In chiral magnets, the spin configurations at the surface can significantly differ from those in the bulk due to the lack of translational symmetry, anisotropy, and surface Dzyaloshinskii-Moriya interaction \cite{tan_giantDMI2024}. These factors may cause surface twists that change the helicity angle of the spin configurations at the surface compared to the bulk \cite{zhang2018direct,Zhang_tomography2018,Jin2023,leonov2016,aqeel2021all,milde2020field}. Recent investigations highlight the surface twists of spirals and skyrmion phases in Cu$_2$OSeO$_3$ \cite{zhang2018direct,Zhang_tomography2018}.
Here, we focus on the spiral and skyrmion phases at the surface of the cubic chiral magnet Cu$_2$OSeO$_3$ that exist at low temperatures by applying a magnetic field along the $\hkl<001>$ crystallographic directions. We investigate these magnetic phases by using resonant elastic X-ray scattering. REXS enables the identification of various magnetic phases through reciprocal phase mapping using element-specific X-ray energies\cite{pollath2019ferromagnetic}. In the case of the B20-type cubic chiral helimagnets, the crystal's small lattice constant relative to the X-ray wavelength leads to the exclusion of structural peaks in diffraction according to Bragg's law. Conversely, in the Cu$_2$OSeO$_3$ crystal with a comparatively large lattice constant \cite{bos2008magnetoelectric} among the B20 members, a forbidden peak emerges in soft X-ray measurements at the Cu $L$-edge.
By employing REXS and characterizing the modulated magnetic orders with the presence of magnetic satellites surrounding the Bragg peaks, one can effectively map magnetic phases in reciprocal space. This comprehensive approach enables a detailed investigation of the magnetic phases of Cu$_2$OSeO$_3$.

Figure~\ref{csophase} highlights the main magnetic phases of a bulk Cu$_2$OSeO$_3$ single crystal when a magnetic field is applied along the $\hkl<001>$ crystallographic direction. Below the critical temperature ${\rm T_c}$, Cu$_2$OSeO$_3$ favors helimagnetic long-range order in a large region of the phase diagram~\cite{Adams2012,seki2012}, characterized by a helical pitch, $\lambda_\text{h}$, and wave vector \qh (Fig. \ref{csophase}(c))\cite{zhang2016resonant}. The balance between magneto-crystalline cubic anisotropy and the anisotropic exchange terms determine the helical spiral orientation along the easy anisotropy direction~\cite{seki2012} which are $\hkl<100>$ and $\hkl<111>$ for Cu$_2$OSeO$_3$ and MnSi, respectively.
 When a finite magnetic field $H$ is applied, it aligns the helix axis and tilts the magnetic moments towards the field direction giving rise to a conical magnetic phase retaining the same pitch $\lambda_\text{h}$ (see Fig.~\ref{csophase}(c),(d)). 

At large applied magnetic fields \(H\ge \rm{H_{c2}}\), all magnetic moments fully align along the applied magnetic field direction, resulting in the establishment of a field-polarized state.
Close to \tc, a skyrmion lattice is observed in a narrow field-temperature region, referred to here as the HTS phase. The HTS is not only observed in Cu$_2$OSeO$_3$ but is also inherent in other cubic helimagnets like Fe$_{1-x}$Co$_x$Si \cite{yu2010real}, MnSi \cite{muhlbauer2009skyrmion} and FeGe \cite{yu2011near}. It is stabilized by thermal fluctuations and as a result observed at elevated temperatures near \tc. Recently, a tilted conical spiral (TC)~\cite{qian2018new} and another skyrmion lattice phase (referred to here as low-temperature skyrmion phase - LTS)~\cite{chacon2018observation} were observed to be stabilized due to cubic crystalline and exchange anisotropy contributions at low temperatures when the external magnetic field is applied along the $\hkl<100>$ crystallographic direction. Note that the LTS is an independent skyrmion lattice phase and compared to thermodynamically stabilized HTS \cite{halder2018thermodynamic,chacon2018observation} it is nearly isotropic for magnetic fields applied along different crystallographic directions.
Furthermore, at relatively high magnetic fields, the spiral orientation deviates from alignment with the magnetic field direction, characterized as a TC phase originating from the competition between cubic anisotropy with $\hkl<100>$ easy axis and exchange anisotropy with $\hkl<111>$ easy axis, specifically when the field is close to \hctwo (Fig. \ref{csophase}(e)) \cite{qian2018new}. 
The transition from a field-polarized state to a topologically non-trivial LTS phase is rather complicated and the system goes through a TC phase before settling in a LTS phase. The LTS and TC phases are strongly hysteretic and depend on the field sweep directions. 

In this work, we experimentally detect an orientationally ordered skyrmion phase by applying a magnetic field along the $\hkl<001>$ direction by following the cycling field protocol at low temperatures. Most interestingly, surface effects lead to the unexpected formation of a robust distorted tilted conical spiral phase (Fig. \ref{csophase}(f)) that persists throughout the field cycling process used to populate skyrmions in the LTS phase.

\section{Experimental} \label{sec:method}
REXS experiments are performed using the ALICE \Romannum{2} endstation, at the PM3 beamline at BESSY \Romannum{2} synchrotron, Berlin.
Cu$_2$OSeO$_3$ single crystals were grown by chemical vapor transport \cite{aqeel2022growth} and then oriented with a Laue diffractometer. One Cu$_2$OSeO$_3$ crystal was precisely cut into a cuboid with 3$\times$2 mm$^2$ lateral dimensions and 1 mm thickness. The crystal was oriented with a surface normal along $\hkl[001]$ and edges along $\hkl<110>$ crystallographic directions and then the top surface was polished mechanically~\cite{aqeel2014surface}.
The sample was mounted on a copper sample holder, with the polished surface facing upward. The sample holder includes a cylindrical temperature shield with a 210\(^\circ)\) cut open arc of 2 mm width that enables a free path for incoming and outgoing X-ray beams.
All REXS results presented in this paper were acquired at the L$_3$ edge using a photon energy of 931.8 eV ($\lambda\approx$ 1.33 nm) and circularly polarized X-rays, which probe the magnetization component parallel to the incident beam.
The forbidden structural Bragg peak at approximately \(2\theta=96.5^\circ\) is observed exclusively in resonant X-ray scattering\cite{zhang2016multidomain,zhang2016resonant}. Resonant X-ray scattering is sensitive to the local point group symmetry, and thus the symmetry of the measurement can sometimes break the general symmetry of the crystal, including features like glide planes or screw axes that usually cause certain reflections to extinct. In the case of Cu$_2$OSeO$_3$, the (001) reflections are typically forbidden by the screw-axis symmetry. Their appearance in REXS measurements can be explained by the anisotropy of the tensor of X-ray susceptibility, which can allow these reflections and their magnetic satellites to become visible. Fortunately, due to the large lattice constant of the crystal (ca. 0.8925 nm \cite{bos2008magnetoelectric}) the (001) reflection is accessible at the Cu L-edge.

The REXS intensity was collected by the 4k CCD detector (GreatEyes GmbH, Germany) located at a distance of 73 cm from the sample \cite{ukleev2022}. The intense (001) Bragg peak was blocked by a beamstop, allowing to observe the magnetic satellites. In this manuscript, all results are obtained at a base temperature of approximately 8 K, utilizing the closed-cycle cryostat (Stinger, ColdEdge Technologies, USA). To improve the signal-to-noise ratio, we averaged ten images, each measured at 20 seconds exposure time, and then subtracted the background at \B=150 mT (field polarized state). We conducted our measurements by tilting the magnetic field from the $\hkl<001>$ crystallographic direction towards the $\hkl<110>$ direction, denoted as angle $\phi$ (figure \ref{measgeo}), with a maximum tilt angle of 11 degrees. This adjustment was performed to enable a detailed study of the distorted tilted conical phases and their behavior under varying magnetic field strengths.

\section{Results} \label{sec:Results}

\begin{figure}[]
    \centering
    \includegraphics[]{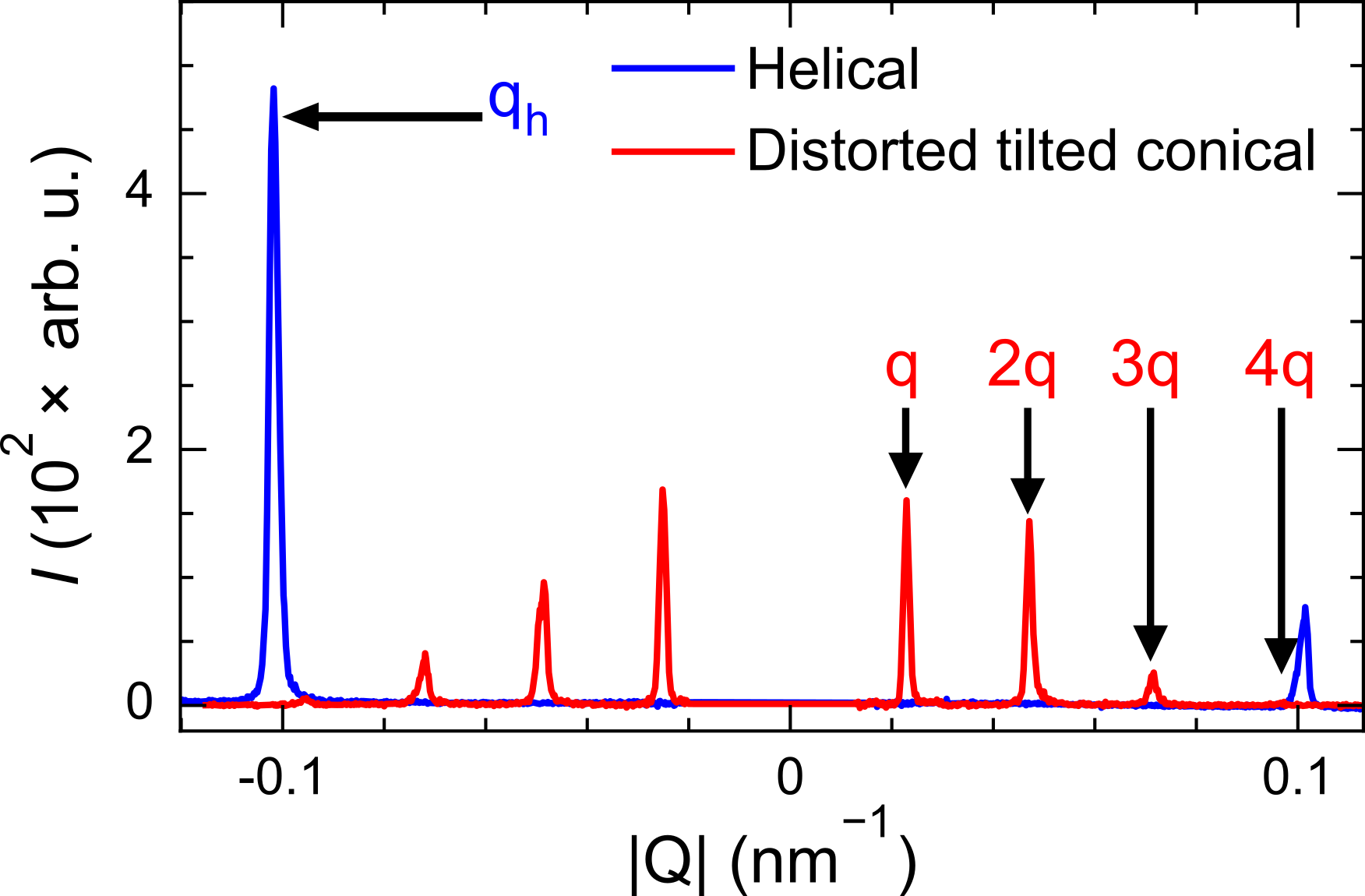}
    \vspace{-0.3cm}
    \caption{Comparison of the REXS intensity profile for helical (blue) and distorted tilted conical spirals (red). The figure highlights a single peak (\qh) from the Friedel pair of the helical spirals at 0~T, and four peaks (q, 2q, 3q, and 4q), each from an equally spaced Friedel pair of the distorted tilted conical at 45~mT, applied at an angle of \(\phi = 6^\circ\) to the $\hkl[001]$ direction. 
    The intensity peak corresponding to the structural Bragg peak is manually removed.}
    \label{intensity-plot}
\end{figure}

\begin{figure*}[ht]
    \includegraphics[]{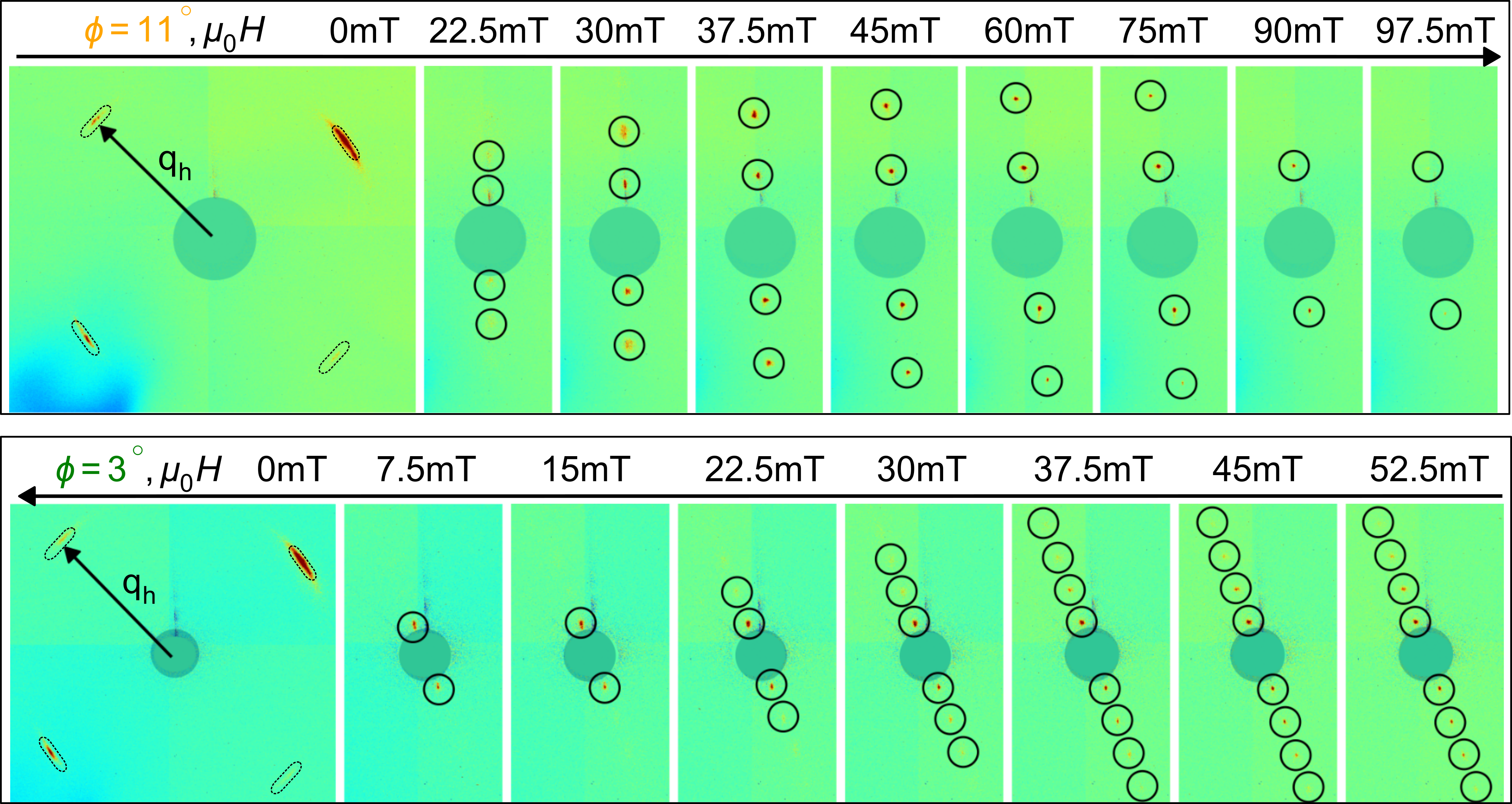}
    \put(-445,220){\text{(a)}}
    \put(-445,95){\text{(b)}}
    \caption{(a) and (b) Evolution of the distorted tilted conical spiral phase under different applied magnetic fields with REXS at 8 K. (a) and (b) show the REXS intensity patterns obtained for two different magnetic field sweep directions. \qh indicates the helical propagation vector defined at 0 T. The circles highlight the positions of higher-order peaks of the distorted tilted conical spiral phase. Note that the measurements shown in (a) and (b) have been carried out for two different magnetic field orientations with respect to the [001] axis represented as $\phi$.}
    \label{evol}
\end{figure*}

\begin{figure}
    \centering
    \includegraphics[]{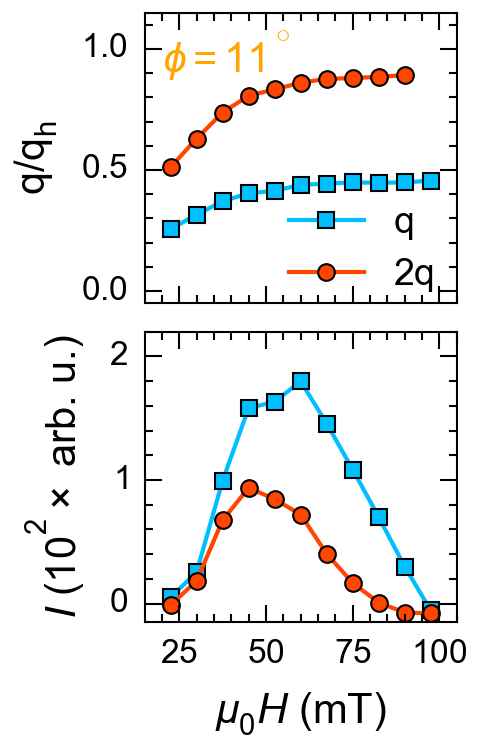}
    \put(-105,180){\text{(a)}}
    \includegraphics[]{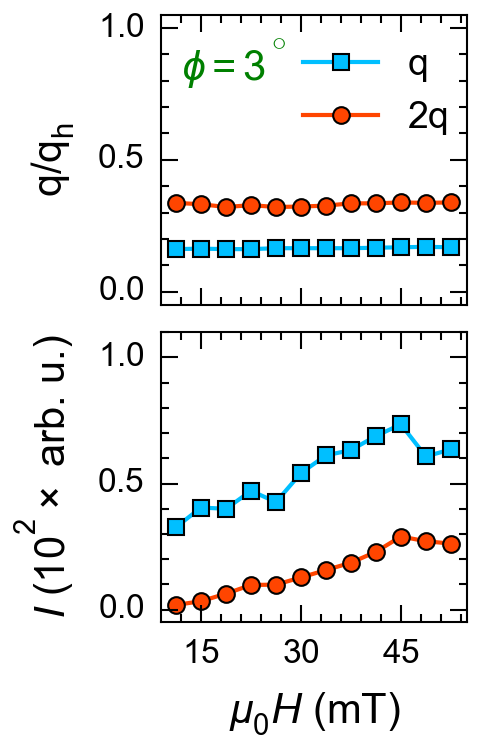}
    \put(-105,180){\text{(b)}}
    \vspace{-0.2cm}
    \caption{(a) and (b) show the change of both propagation vector and intensity of the first and second-order magnetic peaks (q and 2q) as a function of the applied magnetic field for ramping the magnetic field from zero to the field polarized state (a) and ramping the magnetic field to zero (b).}
    \label{evolplot}
\end{figure}

REXS revealed the existence of a periodic anharmonic magnetic order, identified as a distorted tilted conical spiral phase (see Fig. \ref{csophase}(f)), in addition to the proper screw characteristic of the helical spirals in the chiral helimagnetic material Cu$_2$OSeO$_3$.
As shown in Fig. \ref{csophase}(f), the spin texture exhibits a periodic, nonlinear pattern, which deviates from the simple sinusoidal form typically observed in helical structures (Fig. \ref{csophase}(c)). This anharmonic behavior of the spin configuration leads to the appearance of additional higher-order peaks.

Figure~\ref{intensity-plot} shows an exemplary intensity line profile of a CCD image containing the magnetic satellites from REXS in the {\it{hk}} plane of the reciprocal space. The intensity profile shows a single Friedel pair ($\pm q_h$) for the helical spiral at zero applied magnetic fields that are governed by a weak cubic anisotropy \cite{Adams2012}. For the distorted tilted conical phase having a spiral modulation with non-linear spin rotation angles, four Friedel pairs ($\pm$q, $\pm$2q, $\pm$3q, $\pm$4q) are identified at 8 K with \B=45~mT (see Fig.~\ref{intensity-plot}).The propagation vector of the helical spiral, defined as \(q_{\rm h}=2\pi/ \lambda_{\rm h}\), is estimated to be around 0.1 nm$^{-1}$ \cite{seki2012,Adams2012}, yielding a helical spiral wavelength \(\lambda_\text{h}\approx\)~60 nm. However, \qh$\approx4\times$q suggests that the distorted tilted conical spiral phase has a significantly longer wavelength in real space, estimated to be around 240 nm at \(\phi=6^\circ\).

\begin{figure*}[ht]
    \centering
    \includegraphics[]{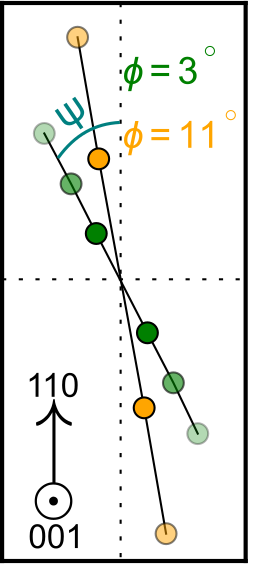}
    \hspace{1cm}
    \put(-65,126){\text{(a)}}
    \includegraphics[]{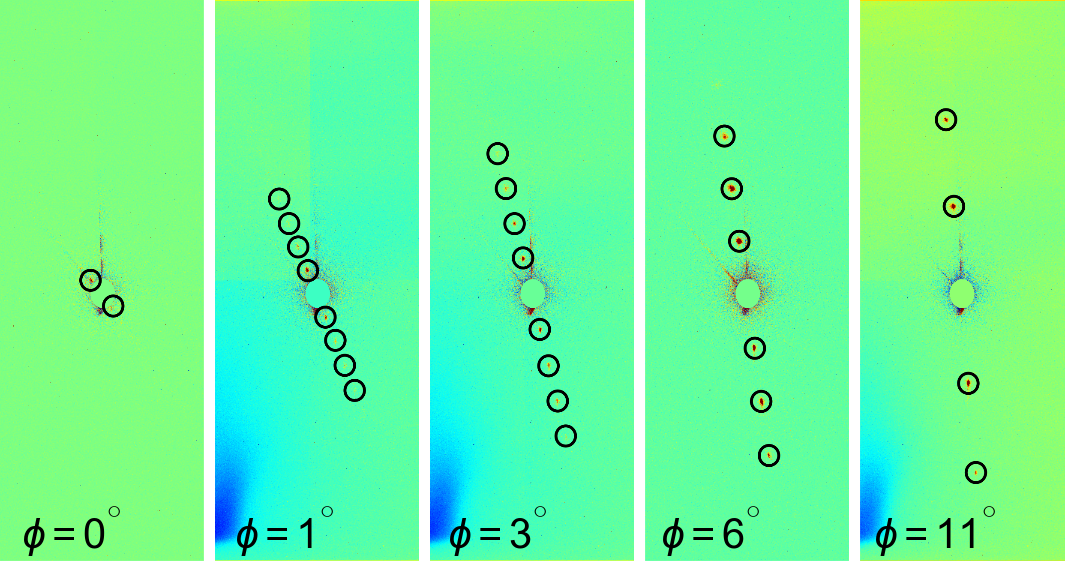}
    \put(-255,126){\text{(b1)}}
    \put(-200,126){\text{(b2)}}
    \put(-150,126){\text{(b3)}}
    \put(-98,126){\text{(b4)}}
    \put(-47,126){\text{(b5)}}
    \includegraphics[]{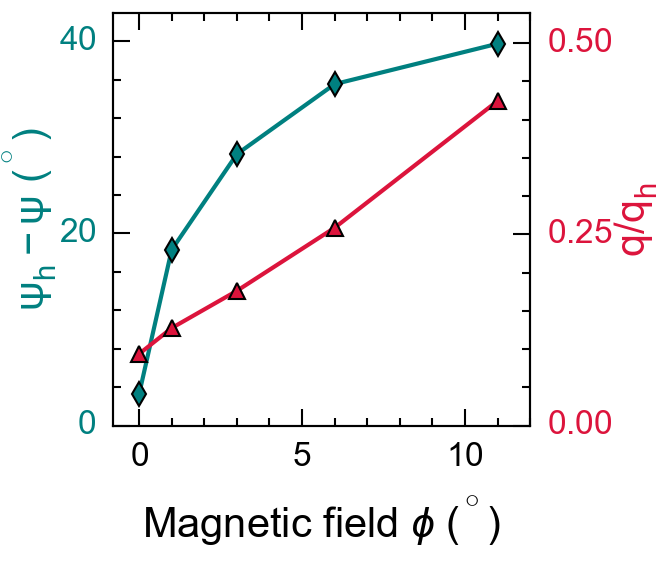}
    \put(-125,120){\text{(c)}}
    \caption{a) Schematics depicting the reorientation of the distorted tilted conical spiral when tilting the applied magnetic field at an angle $\phi$. The reorientation angle $\psi$ is defined for the $\hkl[011]$ in-plane crystallographic direction. b) REXS intensity profile recorded at $T \approx $ 8 K and \B$=$45  mT applied at angle $\phi$. c) Change of reorientation angle ($\psi_h-\psi$) and normalized wave vector (q/q$_h$) of the distorted tilted conical spiral as tilt angle $\phi$ of the applied magnetic field. q$_h$ and $\psi_h$ represent the wave vector and reorientation angle of the helical spiral measured at zero magnetic fields, respectively.}
    \label{angleplot}
\end{figure*}
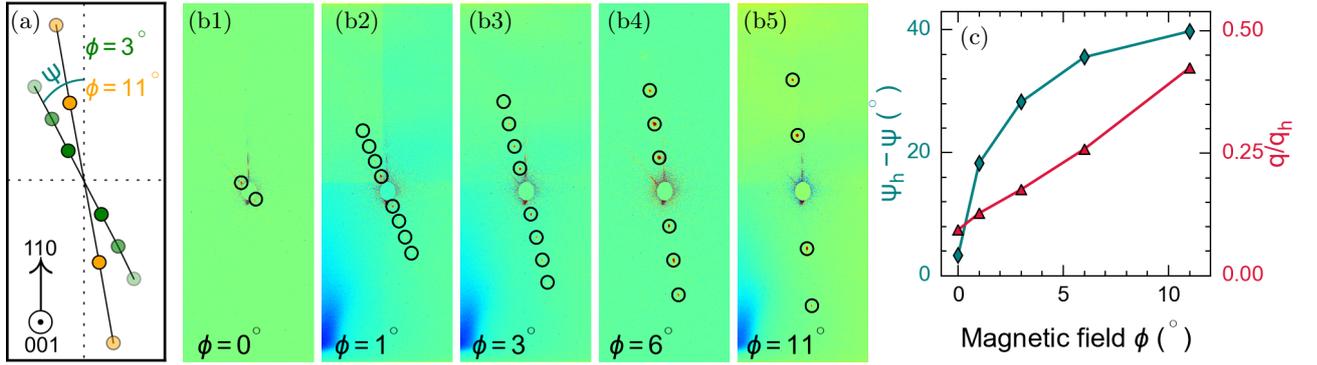

Figures \ref{evol} (a) and  \ref{evol} (b) show the color maps of the REXS intensity in reciprocal space, captured in the {\it{hk}} plane under various applied magnetic fields using a CCD detector for different magnetic field sweep directions.
At zero applied magnetic field, we observe four peaks corresponding to two helical domains, as illustrated in Fig.\ref{evol} (a). With an increase of the magnetic field, the intensity of these helical peaks diminishes, while the distorted TC peaks start to appear at a relatively low magnetic field strength ($\mu_0H \approx 22~ \text{mT}$), as detailed in the supplementary scan [supplementary information]. As the magnetic field is increased further, the wave vector and peak intensity of the distorted TC phase progressively increase until $\mu_0H \leq$ 60 mT, after which a steady decline in peak intensity and saturation of the relative change in wave vector leads to their complete disappearance at about 100 mT, indicating the transition to a field-polarized state, as depicted in Fig. \ref{evolplot} (a,b).  
Note that at smaller magnetic fields close to \hcone applied at an angle $\phi=$11$^\circ$, the distorted TC peaks are fully aligned along the [110] crystallographic direction, as depicted in Fig.\ref{evol} (a). When increasing the applied magnetic field, the distorted TC phase tilts away from the [110] axis with an angle $\psi$. A tilt angle $\psi$ = 4.5~$^\circ$ is observed at 60 mT, as shown in Fig.~\ref{angleplot}, indicating the deviation of the spiral propagation vector towards the easy axis due to the prevailing cubic anisotropy.
A similar trend is observed for an opposite-field sweep direction at $\phi=3^\circ$, with no change in the spiral wave vector, as the magnetic field is decreased from the field-polarized state to 0T (cf. Figs. \ref{evolplot}(a) and \ref{evolplot}(b)).
Importantly, both helical and distorted tilted conical phases were simultaneously observed from 15 mT to 7.5 mT while sweeping the magnetic field to zero (see supplementary information). At zero magnetic field, only four helical peaks are observed.

Recently, higher-order peaks have been observed in a strained Cu$_2$OSeO$_3$  characterized by a chiral soliton lattice with the modulation vector along the strain direction \cite{okamura2017emergence}.
The chiral soliton lattice (CSL) typically forms in uniaxial (strained) chiral magnets, with a significant reduction in the q vector \cite{okamura2017emergence,honda2020topological} when a magnetic field is applied perpendicular to the propagation vector. In contrast, our study reveals distinct behavior: no significant change in the q vector was observed when the field decreased from 50 mT to the helical phase, while an increase in the q vector was noted as the field increased from 0 T to 60 mT. This was accompanied by a consistent rotation in q space, suggesting the stabilization of a distorted tilted conical state. These findings indicate that the observed phase does not correspond to a CSL but rather defines a unique distorted spiral state.
Recently, a distinct surface state resembling the distorted TC phase was reported using SQUID-on-tip microscopy on the surface of a bulk  Cu$_2$OSeO$_3$ crystal. This study identified a stripe surface state and a tilted spiral state in real space, with the stripe state exhibiting higher-order magnetic satellites through FFT analysis of the real-space images~\cite{marchiori2024imaging}.

By tilting the magnetic field, the distorted TC phase reorients, and the number of observed higher-order peaks decreases in the reciprocal map space, as illustrated schematically in Fig. \ref{angleplot}(a).
At $\phi=0^\circ$, the distorted TC phase is oriented along the [010] crystallographic direction, parallel to one of the observed helical domains at 0 T, resulting in \(\rm \psi_h - \psi \approx0\) as shown in  Fig.~\ref{angleplot}(b1). As the tilt angle of the magnetic field, $\phi$, increases the distorted TC phase starts to reorient towards the [110] axis (along the vertical line drawn in the schematics of the CCD image in Fig.~\ref{angleplot}(a)), resulting in an increase of the normalized pitch wave vector (q/q${\rm _h}$) and spiral reorientation angle \(\rm \psi_h - \psi\) (see Fig.~\ref{angleplot}(c)).
Importantly, a substantial increase in the wave vector of the distorted TC phase is observed as a function of applied magnetic field tilt angle $\phi$, which shifts the higher-order peaks outwards compared to the center of the CCD image and reduces the number of observable higher harmonic peaks in the reciprocal space image (cf. Figs. \ref{angleplot}(b1) to (b5)).  
The linear increase in the wave vector (q/q${\rm _h}$) with increasing tilt angle $\phi$ is due to the projection of the distorted TC phase onto the (110) plane. As the spiral tilts away from the magnetic field, it projects more clearly onto this plane at an angle less than 90 degrees.

\begin{figure}[]
    \centering
    \includegraphics[]{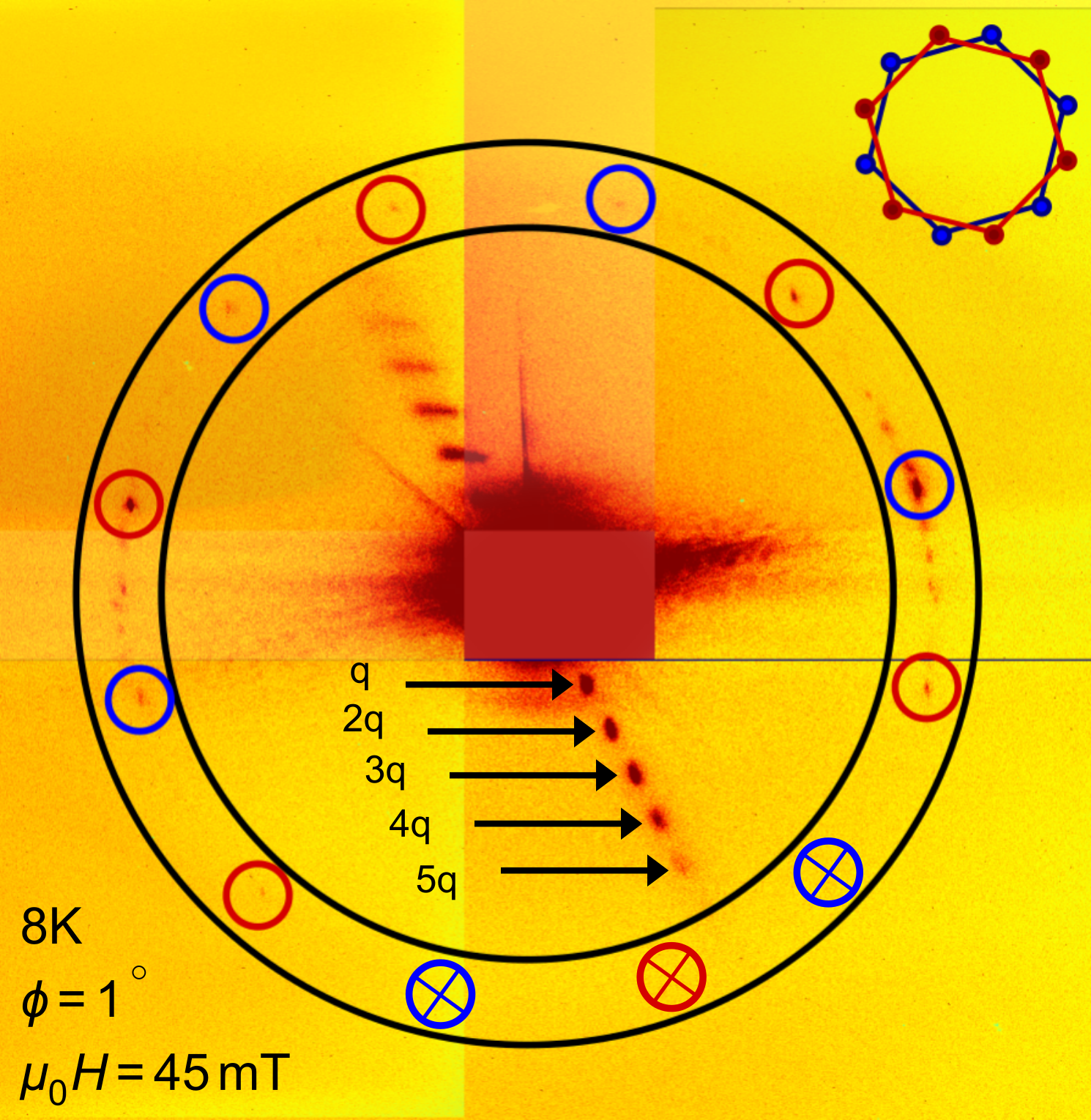}
    \caption{Reciprocal space map elaborating the coexistent distorted tilted conical spiral and skyrmion lattice phases captured through CCD imaging at \B$=$45 mT and field angle $\phi$=1$^\circ$ at 8 K. The red and blue circles highlight the peaks associated with two hexagonal skyrmion lattice domains. The inset shows the schematic diagram of two hexagons (red and blue) that are differently oriented w.r.t each other. Note that due to technical constraints, we did not observe magnetic satellites within the crossed circles.}
    \label{oval}
\end{figure}

 A hexagonal single-domain skyrmion lattice is formed by three coplanar propagation vectors aligned 60\(^\circ\) with respect to each other \cite{zhang2016resonant}. Whereas, within a multi-domain skyrmion lattice, the six-fold symmetric peaks divide into multiple six-fold subsets that are simultaneously sampled by the wide X-ray beam \cite{zhang2016multidomain,langner2014coupled}.

Figure \ref{oval} reveals the reciprocal map of the skyrmion lattice state with six magnetic satellites linked to the structural Bragg peak. It demonstrates the coexistence of both distorted tilted conical spiral and multidomain skyrmion lattice phases at \B~$=$~45~mT. The intensity pattern shown in Fig. \ref{oval} is recorded in four instances with an exposure time of 1200 seconds for each around the Bragg peak to avoid intensity saturation within the CCD and later stitched together to show a full pattern for the skyrmion lattice.
The area bounded by two concentric circles in Fig. \ref{oval} marks the region where we anticipate observing a hexagonal scattering pattern of a skyrmion lattice
\cite{chacon2018observation}. 
The blue and red circles highlight the peaks with higher intensity and are considered as two hexagonal skyrmion lattices which are oriented differently (depicted in the inset) and appear as a multi-domain skyrmion phase.
Moreover, in contrast to the representation in Figure \ref{angleplot} (b) with $\phi=1^\circ$, prolonging the exposure time on the CCD leads to improved visibility of higher harmonic TC phase peaks up to the fifth order, as demonstrated in Figure \ref{oval} with arrows.
When utilizing the REXS technique, its high sensitivity to the sample's surface allows us to precisely identify the magnetic textures present near the surface. This capability highlights unique surface magnetic textures, which may differ from the bulk configuration due to surface anisotropy and surface Dzyaloshinskii-Moriya interaction \cite{tan_giantDMI2024}, leading to the observation of distinctive distorted spin textures that are not present in bulk chiral systems\cite{aqeel2021all, Azhar2022}.

\section{Summary} \label{sec:conclusions}
We experimentally identified an ordered surface texture referred to here as distorted titled conical spiral (TC) phase  
in an extended field region of the magnetic phase diagram at low temperatures. The distorted TC phase points towards the presence of additional magnetic interactions playing an important role at the surface.

The distorted TC phase does not disappear with field cycling; instead, it coexists with the multidomain skyrmion phase populated by the field cycling protocol. The distorted TC and skyrmion phases show a strongly hysteretic behavior depending on the direction of the applied magnetic field sweep. 
Additionally, we examined the reorientation of the distorted TC phase by tilting the static magnetic field relative to the crystallographic direction of Cu$_2$OSeO$_3$. Notably, the distorted TC phase propagation vector follows the magnetic field in the ($\hkl[001]$-$\hkl[011]$) plane with a deviation angle $\psi$. We observed a nonlinear change in angle $\psi-\psi_h$ for magnetic field angle ($\phi$) from almost zero degrees (at $\phi=0^\circ$) to about 40 degrees (at $\phi=11^\circ$) in the {\it{hk}} plane.
Furthermore, we recognized a linear increase of the normalized modulation vector $|q|/|q_h|$ by tilting the magnetic field with angle $\phi$.

Our findings provide valuable insights into the periodicity and orientation of the rarely studied distorted TC phase at the surface of Cu$_2$OSeO$_3$ crystals with a remarkably long period of approximately 240 nm. Resolving the distorted TC phase and other magnetic textures at the surface highlights the importance of REXS for identifying and engineering chiral surface twists. This research opens new avenues for investigating the behavior of these intricate spin textures under various conditions, contributing to the broader knowledge of chiral magnetic materials.

\section{Acknowledgements} \label{sec:acknowledgements}
The authors express their gratitude to C. Pfleiderer, D. Mettus, A. Bauer, K. Everschor-Sitte, and M. Azhar for their valuable discussions. The REXS experiment was conducted at the BESSY II synchrotron (Helmholtz-Zentrum Berlin) as part of proposal 231-11958. This work has been funded by the excellence cluster MCQST under Germany’s Excellence Strategy EXC-2111 (Project No. 390814868). We acknowledge financial support for the VEKMAG project by the German Federal Ministry for Education and Research (BMBF 05K2010, 05K2013, 05K2016, 05K2019) and by HZB. F.R. acknowledges funding by the Deutsche Forschungsgemeinschaft (DFG; German Research Foundation) under SPP2137 Skyrmionics/RA 3570 and A.A. was supported by the DFG grant - 528001743.

%

\newpage

\onecolumngrid
\appendix*

\section{Appendix} \label{Appendix}

\begin{figure}[ht]
    \centering
    \includegraphics[]{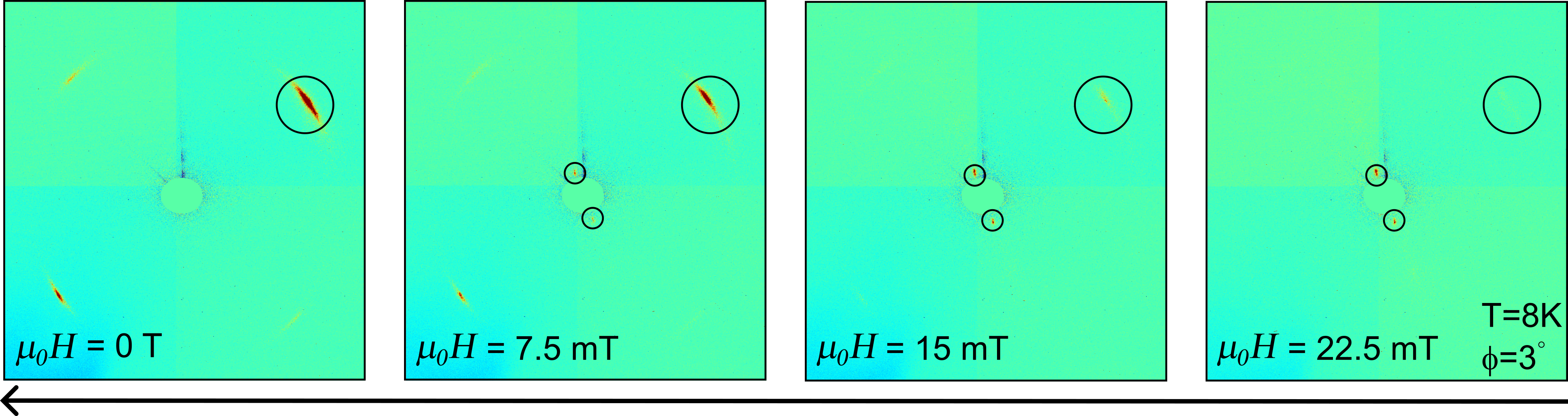}
    \caption{Color maps of REXS intensity in reciprocal space for four different magnetic fields while reducing the field to zero at 8K with $\phi=3^\circ$. The black arrow shows the magnetic field sweeping direction. At 22.5 mT, two peaks associated with the distorted tilted conical are observed. Between 15 mT and 7.5 mT, both distorted tilted conical (enclosed by small circles) and helical (enclosed by a large circle) magnetic satellites are visible. Only the four magnetic satellites corresponding to the helical spiral state are observed at zero magnetic fields.}
    \label{helicalObliqeCoex}
\end{figure}

Figure \ref{helicalObliqeCoex} shows the REXS intensity pattern while the magnetic field is swept to zero at an angle of $\phi=3^\circ$ at 8K. We expect four magnetic satellites in the {\it{hk}} plane corresponding to the helical phase at low magnetic fields. When the magnetic field reaches 22.5 mT, these four peaks start to appear with low intensities, becoming more prominent as the magnetic field decreases to zero (see the large circle in figure \ref{helicalObliqeCoex}). In our study, we observed these helical satellites along with the coexistence of distorted tilted conical (highlighted by small circles) within a narrow magnetic field range from 22.5 mT to 7.5 mT, but only when the magnetic field is swept from field-polarized to zero.

\begin{figure}[h]
    \centering
    \includegraphics[scale=1]{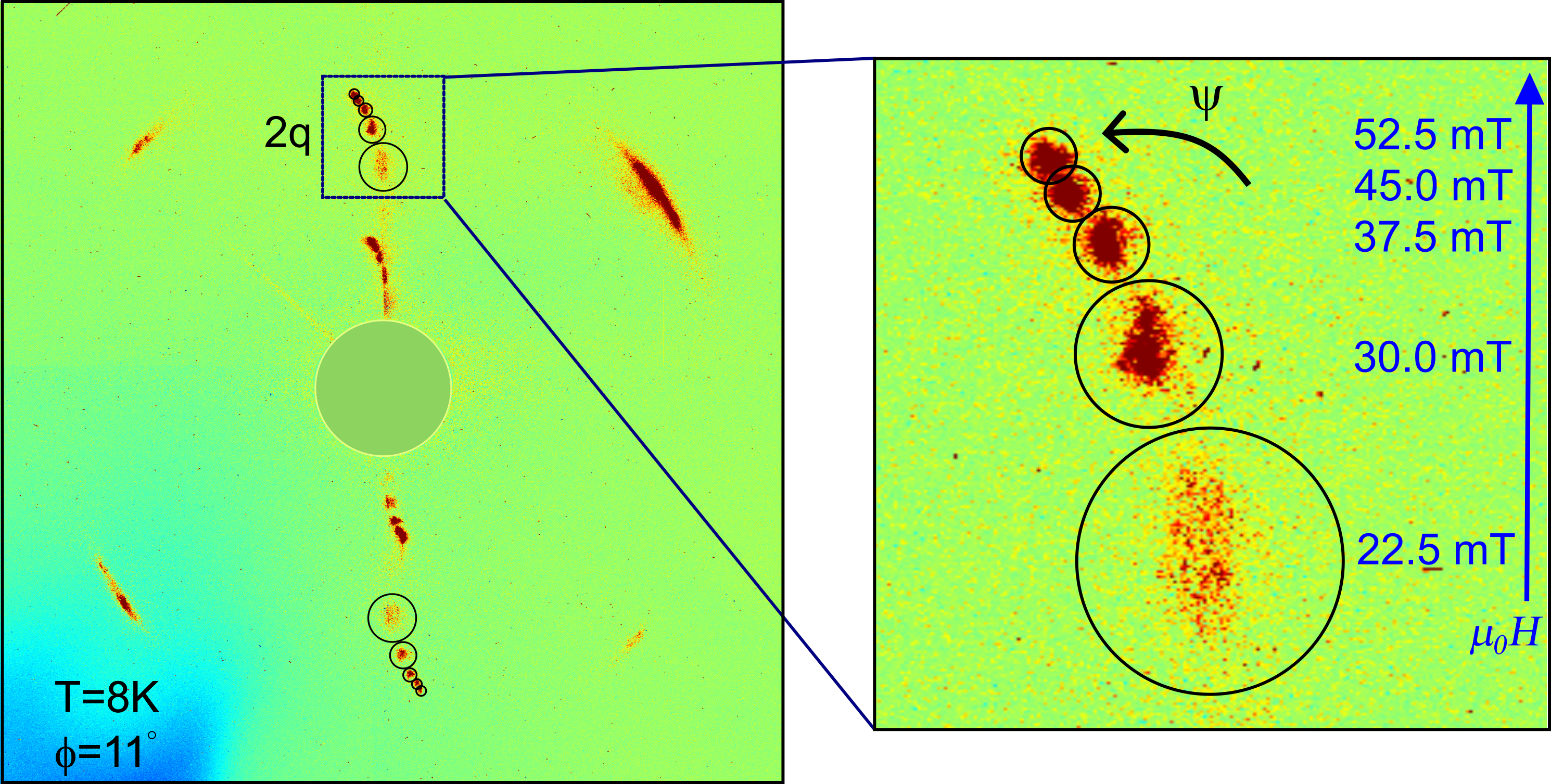}
    \caption{Stacked image of the REXS measurements for the magnetic field sweeping from 0 to 52.5 mT in 7.5 mT increments at $\phi = 11^\circ$. The four magnetic satellites correspond to the helical state, and when the magnetic field reaches 22.5 mT, the first and second harmonic peaks (q and 2q) of the distorted tilted conical phase emerge with a spread intensity pattern (surrounded by a larger circle). As the magnetic field increases, these peaks move outward with respect to the Bragg peak, accompanied by a tilt. The evolution of the second harmonic peaks (2q) indicates the tilting of the spiral propagation vector with increasing magnetic field strength.}
    \label{spiralrotation}
\end{figure}

To study the evolution of the distorted tilted conical phase under varying magnetic field strengths, we compiled REXS intensity patterns from 0 T to 52.5 mT (9 instances), as shown in Figure \ref{evol} of the main manuscript. These patterns allow us to observe how the spiral structures respond to different magnetic field strengths.
Initially, from zero magnetic field up to 15 mT, we observed only the four magnetic satellites associated with the helical state. This indicates that at lower magnetic fields, the system maintains a helical configuration without any noticeable transition to other states.
As the magnetic field increases beyond 22.5 mT, the peak intensities of the distorted tilted conical phase become apparent. This marks a significant transition point where the magnetic field strength is sufficient to change the configuration of the spiral structures. At 22.5 mT, the q and 2q peaks align with the $\hkl<011>$ direction, forming a broad intensity spot highlighted by a large circle in figure \ref{spiralrotation}.
With further increases in the magnetic field, the propagation vector of the distorted tilted conical continues to gradually tilt away from the $\hkl<011>$ direction ($\psi\approx 5^\circ$).
This tilting is accompanied by changes in the modulation q vector, as illustrated in Figure \ref{evolplot} of the main manuscript.
For magnetic fields above 22.5 mT, the peak intensities become narrower, as indicated by the smaller circle in Figure \ref{spiralrotation}. This narrowing suggests that the spiral structures are becoming more defined and possibly more stable under higher magnetic field strengths.
This specific behavior has been observed only when sweeping the magnetic field from zero to a field-polarized state. The process of gradually increasing the magnetic field allows us to capture the transitions and reorientations of the spiral structures in detail.
Moreover, from 52.5 mT to 100 mT, no changes were observed in the modulation vector or the propagation vector of the distorted tilted conical. This plateau indicates that the spiral structures have reached a stable configuration that does not further evolve with increasing magnetic field strength within this range.

\begin{figure}[h]
    \centering
    \includegraphics[scale=1]{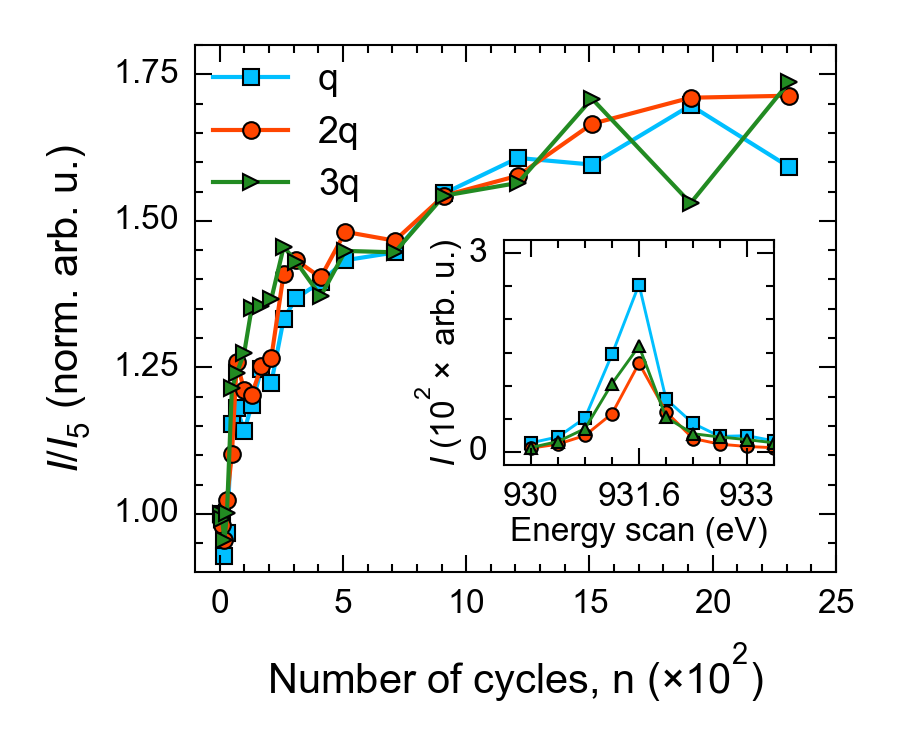}
    \caption{Normalized peak intensity ($I/I_5$) of the peaks in the distorted tilted conical phase as a function of number of cycles $\rm n$ of the static applied magnetic field in the range of 52 mT to 45 mT at  $T \approx$ 8 K. Where $I_5$ represents the peak intensity measured after five field cycles. After field cycles n, each data point is recorded at \B$=$45 mT. The inset shows the intensity amplitude for different energies in the vicinity of L$_3$ edge (931.8 eV) for the first, second, and third higher-order peaks of the distorted tilted conical.}
    \label{cycle}
\end{figure}

Figure \ref{cycle} indicates an increase in the intensity of the peaks observed in the distorted tilted conical phase with magnetic field cycling in the range of \B$_\text{low}=$ 45 mT to \B$_\text{high}=$ 52 mT within the spiral magnetic phase.
The field cycling impact on peak intensities of |q|,|2q|, and |3q| (first, second, and third harmonic, respectively) are shown after normalizing the corresponding intensities with the intensity observed after 5 cycles. The inset in figure \ref{cycle} shows the intensity amplitude of the peaks as a function of X-ray beam energy that confirms the highest intensity occurs for the energy around L$_3$ edge. Notably, the intensity starts at a baseline normalized value of 1 and demonstrates an initial rapid increase, reaching almost 1.5 after 500 cycles. Thereafter, the intensity exhibits a gradual rise, ultimately approaching a saturation point after 1500 cycles. Note that the intensities shown in figure \ref{cycle} do not represent integrated intensity. Therefore, the observed increase in the peak intensity with field cycling can be just a relative increase at the measured spot.

\end{document}